\documentclass[conference]{IEEEtran}
\IEEEoverridecommandlockouts
\usepackage{cite}
\usepackage{amsmath,amssymb,amsfonts}
\usepackage{algorithmic}
\usepackage{graphicx}
\usepackage{float}
\usepackage{caption}
\usepackage{placeins}
\usepackage[table]{xcolor}
\usepackage{textcomp}
\usepackage{subcaption}
\usepackage{mwe}
\usepackage{url}
\usepackage{multirow}
\usepackage{booktabs}
\usepackage{comment}
\usepackage{xcolor}
\usepackage{textcomp}
\usepackage{stfloats}
\usepackage{url}

\def\BibTeX{{\rm B\kern-.05em{\sc i\kern-.025em b}\kern-.08em
    T\kern-.1667em\lower.7ex\hbox{E}\kern-.125emX}}

\begin{document}

\title{Reconstruction of 3D lumbar spine models from incomplete segmentations using landmark detection}

\author{Lara Blomenkamp$^{1}$, Ivanna Kramer$^{1}$, Sabine Bauer$^{3}$, Kevin Weirauch$^{2}$ and Dietrich Paulus$^{1}$
\thanks{*This work was not supported by any organization}
\thanks{*This paper is accepted at 46th Annual International Conference of the IEEE Engineering in Medicine and Biology Society, 2024.}
\thanks{$^{1}$All authors are with the Active Vision Group, Institute for Computational Visualistics, University of Koblenz, Germany. }
\thanks{$^{2}$This author is with the Intelligent Autonomous Systems Group, Institute for Computational Visualistics, University of Koblenz, Germany. }
 \thanks{$^{3}$ This author is with  Institute for Medical Technology and Information Processing, University of Koblenz, Germany. 
    \newline
    {Corresponding email - \tt\small ikramer@uni-koblenz.de}}%

}

\maketitle

\begin{abstract}
Patient-specific 3D spine models serve as a foundation for spinal treatment and surgery planning as well as analysis of loading conditions in biomechanical and biomedical research.
Despite advancements in imaging technologies, the reconstruction of complete 3D spine models often faces challenges due to limitations in imaging modalities such as planar X-Ray and missing certain spinal structures, such as the spinal or transverse processes, in volumetric medical images and resulting segmentations. In this study, we present a novel accurate and time-efficient method to reconstruct complete 3D lumbar spine models from incomplete 3D vertebral bodies obtained from segmented magnetic resonance images (MRI). In our method, we use an affine transformation to align artificial vertebra models with patient-specific incomplete vertebrae. The transformation matrix is derived from vertebra landmarks, which are automatically detected on the vertebra endplates.
The results of our evaluation demonstrate the high accuracy of the performed registration, achieving an average point-to-model distance of 1.95 mm. Additionally, in assessing the morphological properties of the vertebrae and intervertebral characteristics, our method demonstrated a mean absolute error (MAE) of 3.4° in the angles of functional spine units (FSUs), emphasizing its effectiveness in maintaining important spinal features throughout the transformation process of individual vertebrae. Our method achieves the registration of the entire lumbar spine, spanning segments L1 to L5, in just 0.14 seconds, showcasing its time-efficiency. \textbf{Clinical relevance}: the fast and accurate reconstruction of spinal models from incomplete input data such as segmentations provides a foundation for many applications in spine diagnostics, treatment planning, and the development of spinal healthcare solutions.
\end{abstract}


\begin{figure*}[t!]
    \centering
    \includegraphics[width=1\textwidth]{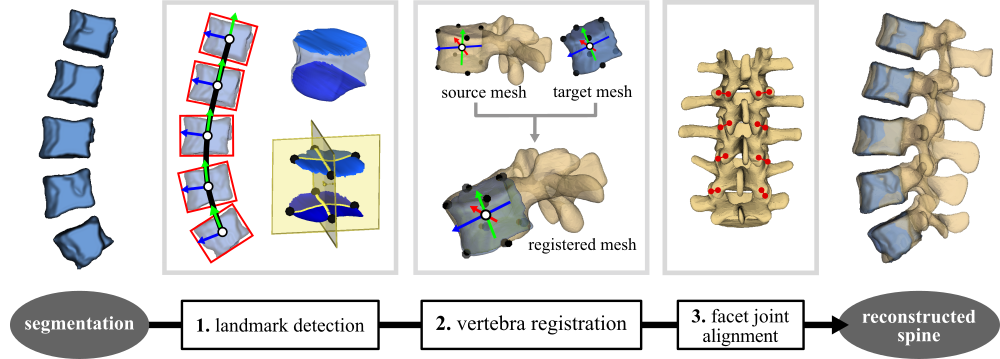}
    \caption{Pipeline of the reconstruction method. Step 1: By determining vertebra orientations and extracting the vertebra endplates, a set of landmarks is determined for each vertebral body. Step 2: From the landmarks, the local coordinate system of the vertebra is calculated and used to determine a transformation for aligning the corresponding vertebra. Step 3: The geometry is adapted to create a coherent spine with realistic facet joint spacing. The resulting spine consists of registered complete vertebrae, that are aligned with the segments of the patient-specific spine.}
    \label{fig:pipeline}
\end{figure*}


\section{Introduction}

3D models of the human lumbar spine are of great interest for analyzing the distribution of forces across spinal structures and their mechanical behaviors. These models are applied in medical research, for diagnosis, treatment planning, and the design and evaluation of spinal implants and medical devices. To create such  models, complete 3D representations of lumbar spinal structures derived from medical imaging data are required. Consequently, there is an increasing demand for automated spine segmentation techniques that can identify and localize the necessary tissues accurately and cost-effectively. However, the challenge with collecting the spinal data is that radiologists and doctors usually focus on specific diseases and only consider certain spine structures when assessing them, leading to a limited consideration of the full range of spinal structures during assessments. Typically, magnetic resonance imaging (MRI) and computed tomography (CT) scans capture only the vertebral bodies and intervertebral discs, often omitting critical structures like the articular processes, transverse processes, and spinous processes. As a result, only the structures that are visible and considered relevant in the imaging data are segmented. Various existing spine segmentation methods
\cite{Zhou2020AVB, Hille2018VBS, Ali2014VBS, Liu2022FLB, Liebl2021ACT, Aslan2012VSI, Stern2010SOV, Kramer2023ASE} focus on the extraction of only the vertebral bodies. In order to reconstruct the whole 3D spine model, incomplete segmentation results require a post-processing step, where the missing parts are completed.

The contributions of this paper are as follows: 1) We propose a novel automated approach to construct a subject-specific 3D lumbar spine model from the incomplete vertebrae extracted from MRI as shown in Fig. \ref{fig:pipeline} by using landmark detection and registration. An essential advantage of our method is its reliance on just eight automatically identified points on the segmented vertebral body. This approach simplifies to the creation of a transformation matrix for registration, enhancing time efficiency and registration accuracy compared to other existing methods. 
2) We make the implementation of our method fully accessible by publishing it as an open-source and ready-to-use 3D Slicer plugin 
\footnote{\url{https://github.com/VisSim-UniKO/3D-Spine-Reconstruction}}, and finally, 3) we publish the ground truth data associated with our study including annotated landmarks and morphological measurements for replication of the results and establishing of new benchmarks.


\section{Related Work}
Various studies have explored the reconstruction of 3D models from volumetric image data.
Benameur et al. \cite{benameur20033d} introduced a method for registration and segmentation of vertebrae using deformable 3D shape models.
The models are used as templates and fitted to radiographic images with a gradient descent method.
Statistical shape models were registered to 3D ultrasound images in a study of Khallagi et al. \cite{khallaghi2010registration}.
Pomero et al. \cite{pomero2004fast} identified anatomical landmarks in biplanar radiographs in order to reconstruct a detailed 3D model.
In the work of Harmouche et al. \cite{harmouche20123d}, an articulated spine model is registered to manually segmented and annotated MRI and X-ray data.
The study by Forsberg et al. \cite{forsberg2013model} employed registration of a spine model to CT image data with an affine transformation based on vertebral rotation estimation, and an deformable phase-based registration algorithm.
The resulting alignment can be refined with an atlas-based non-rigid registration step \cite{forsberg2015atlas}.
Rasoulian et al. \cite{rasoulian2013lumbar} conducted an analysis of the spines shape and pose from a CT image to create an statistical multi-vertebra shape+pose model.
The models are aligned with an expectation maximization registration technique in order to segment vertebrae in the CT image. A study of Kramer et al. \cite{Kramer2023ASE} focused on the creation of the lumbar spine models with patient-specific lordosis using a biomechanical spine model.  Initially, the lordosis curve was extracted from 3D spinal meshes derived from automatically segmented  MR images. Subsequently, this curve was used as a reference for the desired lordosis. The initial spine model was then aligned with the extracted lordosis curve by simulating specific forces, ensuring the model closely mimics the patient's natural spinal curvature.

Despite various proposed techniques for constructing 3D spine models, current research still needs an efficient and fully automated method, where the provided image data is limited to the vertebral bodies.


\section{Materials and Methods}
Our proposed method aims to construct a comprehensive 3D model of the human lumbar spine, addressing the challenges of incomplete segmentation from medical imaging. Often, vertebra segmentation primarily captures a series of vertebral bodies without encompassing the entire vertebrae. A sample 3D model of the segmented vertebral bodies is shown in Fig.~\ref{fig:pipeline}.

In the present study the segmentations are extracted from MRI scans utilizing a 2D U-Net as described in \cite{Kramer2023ASE}. Our approach enhances the incomplete segmented vertebral bodies (VBs) by substituting them with complete vertebra models. For building a target lumbar spine, we used detailed artificial Sawbone vertebra meshes, which represent the average spine anatomy of a European male \cite{wataname2018}.

The proposed method's main focus is to find transformation parameters that match complete vertebra meshes, called source meshes, with the patient-specific incomplete meshes of VBs, called target meshes. 
We determine these parameters by step-wise analyzing the morphological and rotational characteristics of the VBs.
As shown in step 1 in Fig. \ref{fig:pipeline}, key anatomical landmarks are identified first on the meshes. 
In the second step, we use these landmarks to compute the parameters of two transformation matrices, that encapsulate translation and rotation on one hand and scaling on the other hand (see Equation~\ref{eq:matrix}). 
Finally in the last phase, we apply elastic transformations to ensure the precise alignment of the facet joints, preventing any potential collisions of these surfaces. 
 
In the following section, each step of the proposed pipeline is explained in detail. 

\subsection{Landmark detection}
In order to calculate the transformation parameters, we analyse the shape of the vertebral bodies of the source meshes and the target meshes. For each vertebra, we automatically derive 8 landmarks from the posterior, anterior, left and right edge of each VBs endplate as depicted in Fig.~\ref{fig:pipeline} (Step 1), by adopting the method proposed by Kramer et al in \cite{kramer2023VertebraeMeasurements}. 
In their approach, the local anatomical vertebra axes are approximated from an object-oriented bounding box and the derivative of a cubic spline fitted through the vertebrae's centers of mass.
The vertebra endplates are extracted by removing polygons with a low similarity between the surface normal and the calculated longitudinal axis of the vertebra.
In this study, we improve the method for endplate extraction by applying a connectivity filter to the extracted meshes.
This eliminates potential false positive polygons, by removing small regions that are not connected to the endplate's surface.
The required landmarks are determined by identifying the outermost points from the intersection of the extracted endplates with the local anatomical planes of the vertebral meshes \cite{kramer2023VertebraeMeasurements}.


\subsection{Vertebra registration}
\begin{figure}[b!]
    \centering
        \begin{subfigure}{0.23\textwidth}  
        \centering 
        \includegraphics[width=\textwidth]{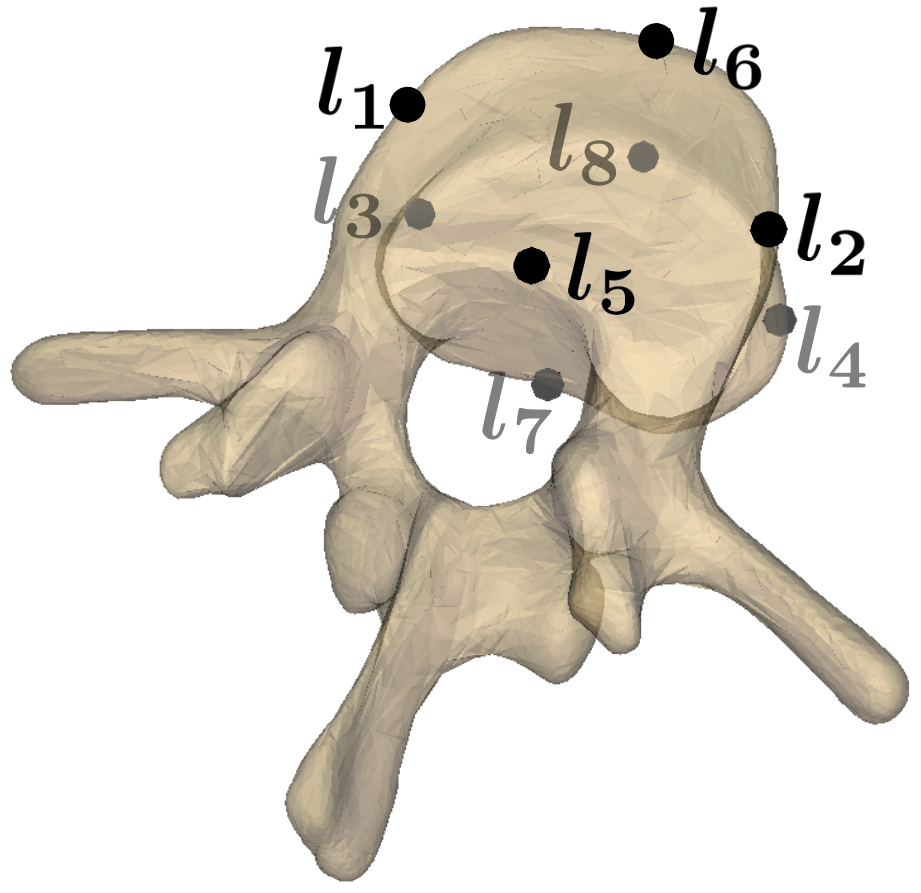}
        \caption{}
    \end{subfigure}
    \begin{subfigure}{0.23\textwidth}   
        \centering 
        \includegraphics[width=\textwidth]{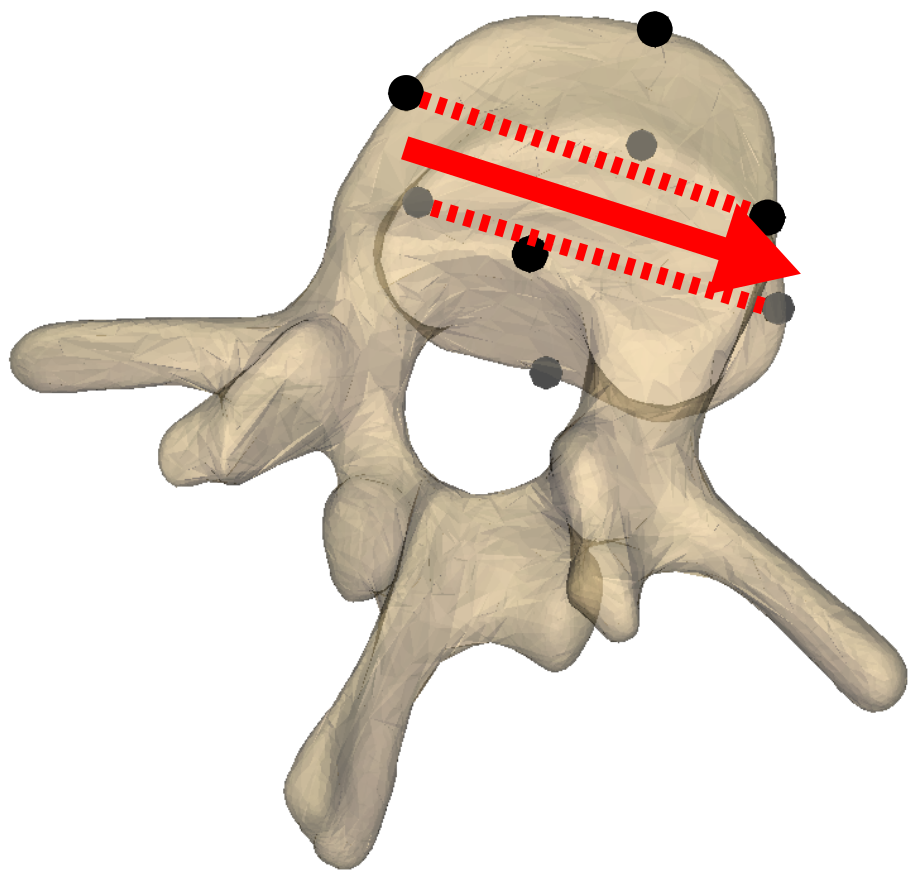}
        \caption{}
    \end{subfigure}
    \begin{subfigure}{0.23\textwidth}   
        \centering 
        \includegraphics[width=\textwidth]{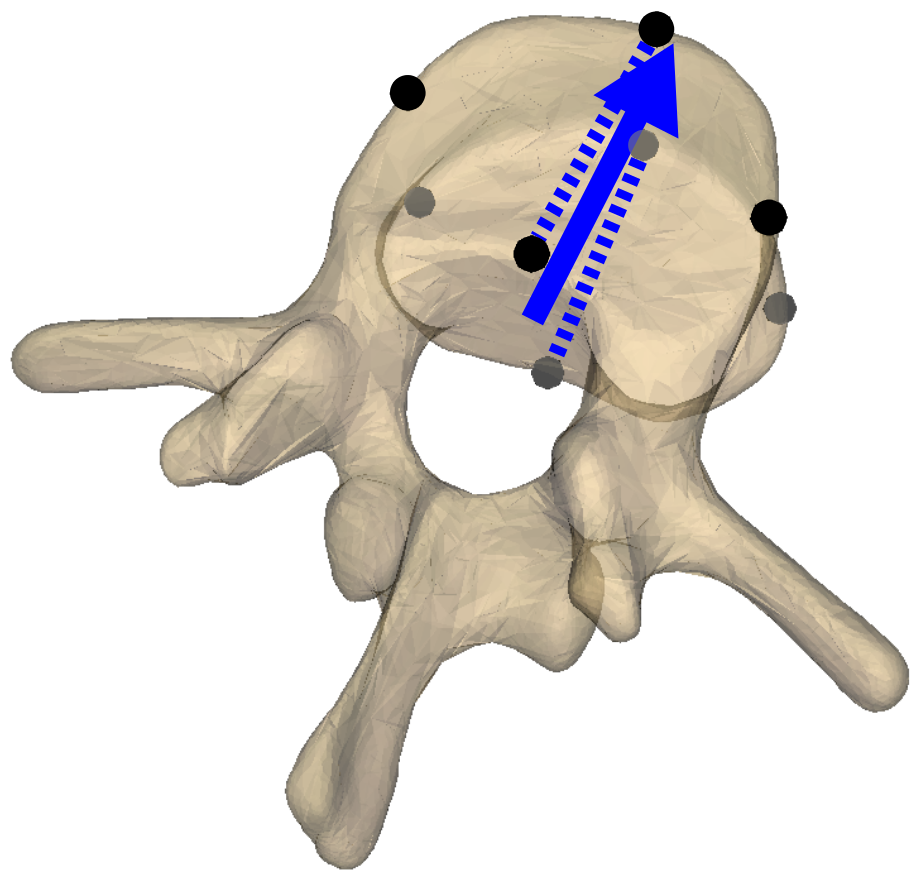}
        \caption{}
    \end{subfigure}
    \begin{subfigure}{0.23\textwidth}   
        \centering 
        \includegraphics[width=\textwidth]{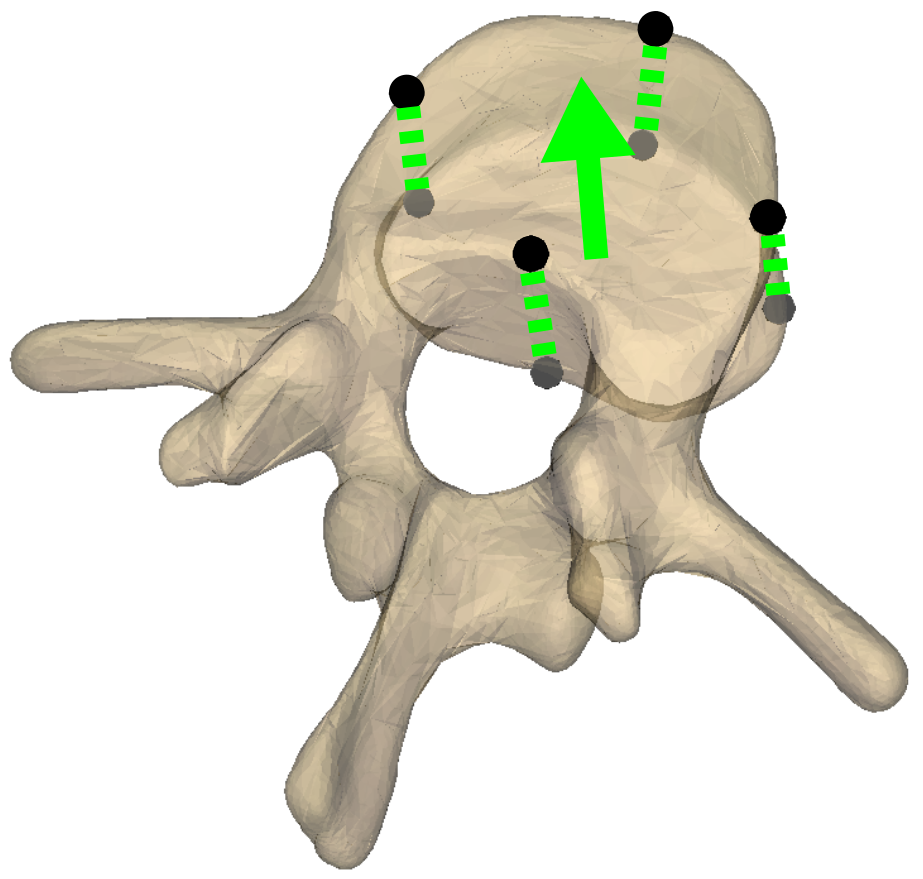}
        \caption{}
    \end{subfigure}
    \caption{(a) A set of key anatomical landmarks is determined for the vertebral body. The vertebra orientation vectors are defined for three local anatomical orientations by averaging vectors between the landmarks: (b) $\vec{x}_g$ in left-right orientation, (c) $\vec{y}_g$ in posterior-anterior orientation, (d) $\vec{z}_g$ in inferior-superior orientation.} 
    \label{fig:dimensions}
\end{figure}

\begin{figure}[htb]
    \centering
    \includegraphics[width=0.5\textwidth]{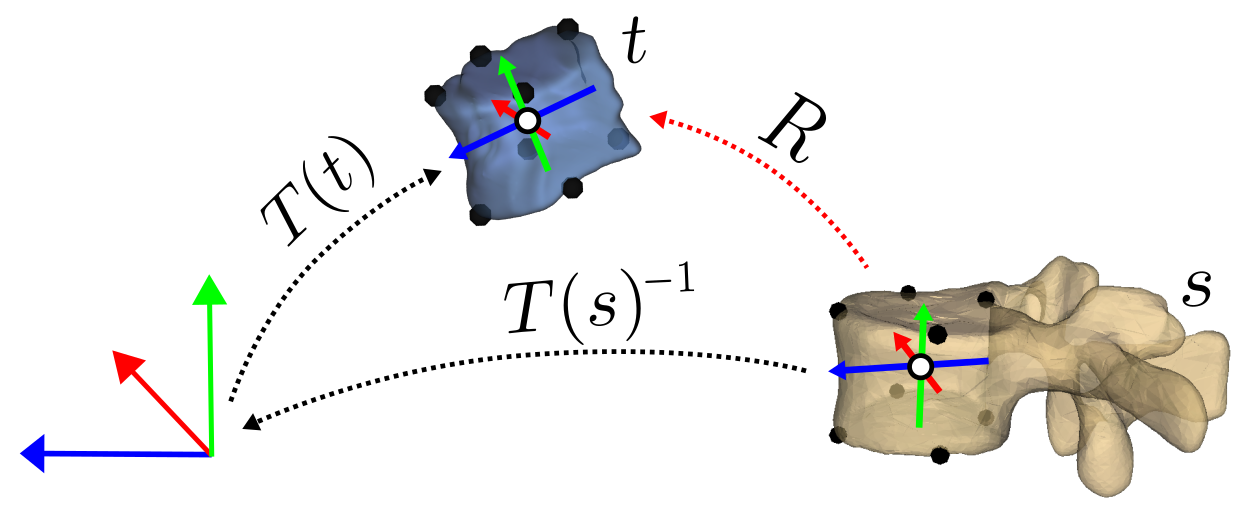}
    \caption{The registration matrix $R$ represents a stepwise transformation from the source mesh $s$ to the global coordinate system and then to the target mesh $t$.}
    \label{fig:registration}
\end{figure}

In order to transform the source mesh~$s$ to the target mesh~$t$, we first need to calculate the transformation $T(s)^{-1}$ for aligning the source mesh with the global coordinate system and then align it with the target mesh by performing a transformation $T(t)$ (see Fig. \ref{fig:registration}). We derive the transformation parameters for both source and target vertebra geometries using the detected landmarks. 

We estimate local orientation, position and scale of each vertebra geometry $g$. The three basis vectors $\vec{x}_g, \vec{y}_g, \vec{z}_g \in \mathbb{R}^{3}$ of the corresponding object coordinate system  (see Fig.~\ref{fig:dimensions}) are calculated 
as follows:

\begin{align}
    \begin{split}
        \vec{x}_g &=
        \frac{1}{2}
        (\overrightarrow{l_1 l_2} + \overrightarrow{l_3 l_4}), \\
        \vec{y}_g &=
        \frac{1}{2}
        (\overrightarrow{l_5 l_6} + \overrightarrow{l_7 l_8}), \\
        \vec{z}_g &=
        \frac{1}{4}
        (\overrightarrow{l_3 l_1} + \overrightarrow{l_4 l_2} + \overrightarrow{l_7 l_5} + \overrightarrow{l_8 l_6}).
    \end{split}
    \label{eq:basis}
\end{align}

The 3D position of the vertebra is determined through the center of mass $\vec{c}_g \in \mathbb{R}^{3}$ of the landmarks. The local scales of the vertebral mesh are defined as magnitudes $\left|\vec{x}_g\right|, \left|\vec{y}_g\right|,\left|\vec{z}_g\right| \in \mathbb{R}$ of the corresponding basis vectors.

Using the transformation parameters, we determine a transformation matrix $T(g)$, describing the vertebra-local alignment in the global coordinate system through translation, rotation and scaling:

\begin{equation}
    T(g) = 
    \begin{bmatrix}
    \hat{x}_g & \hat{y}_g & \hat{z}_g & \vec{c}_g \\
    0 & 0 & 0 & 1 \\
    \end{bmatrix}
    \cdot
    \operatorname{diag}
    (
    \left|\vec{x}_g\right|,
    \left|\vec{y}_g\right|,
    \left|\vec{z}_g\right|,
    1
    ),
    \label{eq:matrix}
\end{equation}
where the rotation parameters $\hat{x}_g, \hat{y}_g, \hat{z}_g \in \mathbb{R}^{3}$ are the normalized versions of $\vec{x}_g, \vec{y}_g, \vec{z}_g$.

To register the source mesh $s$ to the target mesh $t$ (see Fig.~\ref{fig:registration}), the transformation matrix $R$ is calculated:

\begin{equation}
    R = T(t) \cdot T(s)^{-1}.
\end{equation}

Applying this transformation to all target meshes yields a completed custom spine (see reconstructed spine from Fig. \ref{fig:pipeline}).

\subsection{Alignment of facet joints}
To ensure a compatible assembly of the vertebral processes, a method for facet joint alignment \cite{Blomenkamp2023EGO} is deployed.
In this step, the geometry of each facet is warped with an elastic transformation.
This adjusts the geometry of the articular processes to prevent them from both overlapping and excessive spacing (see Fig.~\ref{fig:pipeline},~Step 3).
Consequently, the individual vertebrae are assembled as a coherent spine model.
The required measurements for the facet joint space width were taken from \cite{simon2012vivo} for the lumbar vertebrae. 


\section{Experiments and Results}
To evaluate the proposed method, we used segmentations from a VerSe 2021 dataset published in \cite{sekuboyina2021verse}, including 50 lumbar vertebrae from 10 different spines, as ground truth meshes. We excluded the posterior part of the vertebrae, focusing exclusively on the vertebral bodies as the input for our method. By applying the proposed reconstruction method to the vertebral bodies from the VerSe dataset, we compared the registration outcomes against the original segmentation models. 
To evaluate the effectiveness of our approach relative to established baseline methods, we conducted a series of comparisons. This involved executing the registration process through various methods, integrating the Iterative Closest Point (ICP) registration \cite{besl1992method} within our method, alongside standalone ICP registration, and registration employing ALPACA \cite{Porto2021ALP}. 
We conducted two distinct experiments employing the standalone ICP method. In the first experiment, we aligned artificial vertebrae (source) with patient-specific VBs. For the second experiment, we manually modified the source meshes by removing the posterior part, leaving only the vertebral body for registration. We then computed a rigid transformation matrix between the source and target VBs using ICP. After determining the transformation matrix, we used it to align the complete source mesh with the target VB. In a comparable experiment, we utilized ALPACA to derive an affine transformation matrix. Once obtained, we applied it to align the source vertebra with the target VB.

The registration accuracy was initially assessed by measuring the average point-to-model distance. This metric calculates the distance from each point on the source mesh to the nearest point on the surface of the target mesh for every corresponding pair of registered vertebrae. 
\begin{figure}
    \centering
    \begin{subfigure}{0.46\textwidth}  
        \centering 
        \includegraphics[width=\textwidth]{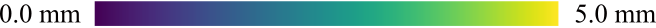}
        \vspace{0.5mm}
    \end{subfigure}
    \begin{subfigure}{0.156\textwidth}  
        \centering 
        \includegraphics[width=\textwidth]{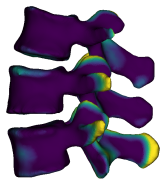}
        \caption{}
    \end{subfigure}
    \begin{subfigure}{0.156\textwidth}   
        \centering 
        \includegraphics[width=\textwidth]{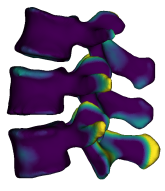}
        \caption{}
    \end{subfigure}
    \begin{subfigure}{0.156\textwidth}   
        \centering 
        \includegraphics[width=\textwidth]{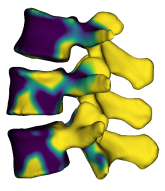}
        \caption{}
    \end{subfigure}
    \begin{subfigure}{0.156\textwidth}   
        \centering 
        \includegraphics[width=\textwidth]{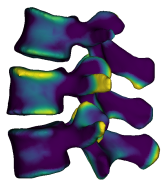}
        \caption{}
    \end{subfigure}
    \begin{subfigure}{0.156\textwidth}   
        \centering 
        \includegraphics[width=\textwidth]{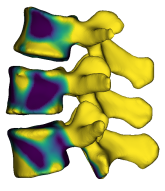}
        \caption{}
    \end{subfigure}
    \caption{Comparison of model surface distances for different registration methods: (a)~Our method, (b)~Our method + ICP, (c)~ICP, (d)~ICP on VB's, (e)~ALPACA on VB's. The custom vertebra models are shown with the color reflecting the distances from each mesh point to the closest surface point of the registered vertebra.} 
    \label{fig:model-distances}
\end{figure}

Since landmarks are determined only from vertebral bodies, the point-to-model distance was measured for both the VBs and the entire vertebra models. The distances from the registered source and target vertebrae are illustrated in Fig.~\ref{fig:model-distances}.

In addition to registration accuracy, we evaluated various morphological features of the vertebrae relevant in clinical contexts, including landmark positions, dimensions of the vertebral body (height, width, depth), the height of the intervertebral disc (IVD), and the angle of functional spinal units (FSUs). The ground truth for these measurements was established through a hybrid approach: morphological features were initially measured automatically using the method described by Kramer et al. in \cite{kramer2023VertebraeMeasurements}, and subsequently refined manually by an expert in the field. We leveraged the aligned landmarks, adjusted through the application of the obtained registration matrix, to calculate the morphological characteristics of the registered vertebrae. To quantify the error between the ground truth and the calculated morphological parameters, we employed the mean absolute error (MAE) metric. 
As an additional  metric, we measured and compared the time required to register the source vertebrae to the patient-specific target models averaged over 50 vertebrae. 

\begin{table*}[h!]
\caption{Average error values of geometric and morphological features for different registration methods.}

\begin{center}
\begin{tabular*}{\textwidth}{lccccccc}
\toprule
\multicolumn{1}{c}{} & \multicolumn{2}{c}{\textbf{Mean Point-to-Model Distance}} & \multicolumn{4}{c}{ \textbf{Morphology (MAE)}} & \multirow{2}{*}{\hspace{-0,8em}\textbf{Time (s)}} \\
\cmidrule(rl){2-3} \cmidrule(rl){4-7}
\textbf{Method} & \textbf{VB (mm)} & \textbf{Vertebra (mm)} & \textbf{Landmarks (mm)}  &  \textbf{Dimensions (mm)} & \textbf{IVD height (mm)} & \textbf{FSU angle (\textdegree)}  \\
\midrule
\textbf{Our}         
    &    1.38
    &    1.95
    &    \textbf{1.36}
    &    \textbf{1.65}
    &    \textbf{0.87}
    &    \textbf{3.4}
    &     \textbf{0.14}\\
\textbf{Our + ICP}
    &    \textbf{1.28}
    &    \textbf{1.84}
    &    1.91
    &    1.76
    &    1.0
    &    3.76
    &   1.0\\
\textbf{ICP}
    &    3.45
    &    6.42
    &    14.24
    &    3.64
    &    16.17
    &    40.58
    &    1.69\\
\textbf{ICP on VBs}
    &    1.99
    &    2.51
    &    3.76
    &    3.64
    &    7.08
    &    25.02
    &  1.37\\
\textbf{ALPACA \cite{Porto2021ALP} on VBs}
    &    5.85
    &    6.29
    &    19.33
    &    4.42
    &    42.20
    &    77.20
    &   83.5\\
\bottomrule
\vspace{0.1cm}
\end{tabular*}
\label{tab:results_evaluation}
\end{center}
\end{table*}


Table~\ref{tab:results_evaluation} presents a comparison of different registration methods based on their accuracy in aligning one model to another and preserving the morphological features of the target model. The comparison reveals, that although our method slightly underperforms in point-to-model distance measurements on vertebral bodies and whole vertebrae compared to its enhancement with the ICP algorithm ("Our + ICP") with an error increase of 0.1 mm, it significantly surpasses other baseline methods with an error margin up to 4.34 mm. Particularly, our method shows the lowest MAE value of 1.36 mm, underscoring its accuracy in landmark registration. In contrast, the  ICP and ALPACA indicate the highest errors of 14.24 mm and 19.33 mm respectively,  when transforming the landmarks. Across various morphological measurements, our approach consistently presents the minimal errors, demonstrating its capability in preservation of the spinal morphological features.

Furthermore, in terms of computational efficiency, our method and its ICP-enhanced version emerge as the fastest, with completion times of 0.14 seconds and 1.0 seconds for model registration, respectively. This efficiency emphasizes the suitability of our method for integration into clinical routine, offering a balance of accuracy and computational efficiency.
All experiments were done on an Intel machine of 11th Gen Intel® Core™ i7-11800H @ 2.30GHz × 16 and 16 GB of RAM.




\section{Conclusion}
In this paper, we proposed a novel approach for reconstructing 3D spinal models from incomplete spinal segmentations, specifically focusing on segmentations that only include the vertebral bodies. The key feature of our methodology lies in the registration of vertebral models using just 8 automatically detected landmark points on each vertebra. This allows for the transformation parameters to be calculated analytically in a non-iterative manner, leading to fast computations,  which is especially important for clinical and research environments. The experiments and the comparison with other registration methods show that the integration of our method with ICP results in the highest registration accuracy, if only with a slight improvement over our method standalone.

In our proposed method, we use artificial Sawbone meshes to align with patient-specific models of incomplete vertebrae, serving as a step towards achieving our ultimate goal of creating accurate complete personalized spinal models. Our future research aims to validate and enhance our proposed method, focusing on its ability to align spinal meshes from diverse imaging sources, such as X-Rays, camera-based systems, that can provide estimated vertebral bodies from the patient in standing and in dynamic conditions with the 3D vertebral models of the same patient gained from accurate CT or MRI scans in supine body position. This effort is important when addressing orthopedic and spinal healthcare challenges by integrating and harmonizing data across different diagnostic tools.

\section{Compliance with Ethical Standards}
This research study was conducted retrospectively using human subject data made available in open access. Ethical approval was *not* required as confirmed by the license attached with the open access data.

\bibliographystyle{IEEEtran}
\bibliography{ref}

\end{document}